\documentclass[12pt]{article} 
\usepackage[pdftex]{graphicx}
\usepackage{amsmath,amssymb,amsthm}
\usepackage{doi}
\usepackage{url}
\usepackage{lettrine}
\usepackage{hyperref}
\usepackage{cleveref}
\usepackage{amsfonts}
\usepackage{pstricks,pst-node,pst-text,pst-3d}
\usepackage{url}
\usepackage{setspace}

\input colordvi.sty %\usepackage[active]{srcltx}
\DeclareSymbolFontAlphabet{\mathbb}{AMSb}

\begin{document}

\begin{center}
\Large{{\bf B\"acklund transformations: a tool to study Abelian and non-Abelian nonlinear evolution equations
}}
\end{center}
\normalsize 
\begin{center}{
{\bf\large Sandra Carillo}
\\{\rm Dipartimento Scienze di Base e Applicate
    per l'Ingegneria \\  {Sapienza}   Universit\`a di Roma,  16, Via A. Scarpa, 00161 Rome, Italy} \\ 
   {\small and} \\ {\rm Gr. Roma1, IV - Mathematical Methods in NonLinear Physics\\ National Institute for Nuclear Physics (I.N.F.N.), Rome, Italy \\ \medskip
   {\bf\large Cornelia Schiebold}
\\{\rm Department of Natural Sciences, Engineering, and Mathematics,\\ Mid Sweden University, Sundsvall, Sweden. \\
%% Cornelia.Schiebold\symbol{64}miun.se}
   }}}
 \end{center}
\medskip%-------------------------------------------------------------------------------
\begin{abstract} %A {\it novel} {\red{ }}
The KdV eigenfunction equation is considered: some explicit solutions are constructed. These, to the best  of the authors' knowledge,  new solutions represent an example of the powerfulness of the method devised. 
Specifically,  B\"acklund transformation are applied to reveal algebraic properties enjoyed by nonlinear  evolution equations they connect. 
Indeed,   B\"acklund transformations, well known to represent a key tool in the study  of nonlinear  evolution equations, are shown to allow the construction of a net of nonlinear links, termed {\it B\"acklund chart}, connecting 
Abelian as well as non Abelian   equations. The present study concerns
third order nonlinear  evolution equations which are all connected to the KdV equation. In particular, 
the Abelian 
 wide B\"acklund chart connecting these nonlinear  evolution equations is recalled.
Then, the links, originally established in the case of Abelian 
equations,  are shown to conserve their validity when
non Abelian counterparts are considered. In addition, the non-commutative case reveals
a richer structure related to the multiplicity of  non-Abelian equations which correspond to the same Abelian one.  Reduction from the nc to the commutative case allow to show  the connection of 
the KdV equation with KdV eigenfunction equation, in the  {\it scalar} case. 
 	Finally, recently obtained
 matrix solutions of the mKdV equations are recalled.							
\end{abstract}

\noindent

\textbf{\it{Keywords:}} % \Keywords
{Nonlinear Evolution Equations;  B\"acklund Transformations;   Recursion Operators; 
Korteweg deVries-type equations; Invariances; Cole-Hopf Transformations.}                                                                                                                                                                                       

{\bf AMS Classification}: {58G37; 35Q53; 58F07} % e.g. 35A30; 81Q05
\section{Introduction}
\label{introd}
The crucial role played by B\"acklund transformation in investigating {\it soliton equations} is well known as testified by  \cite{Ablowitz, CalogeroDegasperis, RogersShadwick, RogersAmes, RogersSchief, Gu-book} referring only to  some well known books on the subject.  Third order nonlinear evolution equations, termed KdV-type, are  studied. Indeed, both in the Abelian case  \cite{BS1, apnum} as well as in the non-Abelian one  \cite{Carillo:Schiebold:JMP2009,SIGMA2016, JMP2018, EECT2019}, a wide net of nonlinear evolution equations  turns out to be connected to the Korteweg-de Vries (KdV) equation or, respectively, to its nc counterpart. The net of links, termed B\"acklund chart, allows to reveal many interesting properties enjoyed by the nonlinear evolution equations it connects. Specifically, all algebraic properties which are preserved under B\"acklund transformations can be transferred from one equation to all the others within the B\"acklund chart. The nonlinear evolution equations under investigation, further to the KdV equation are the potential Korteweg-de Vries (pKdV), the  modified Korteweg-de Vries (mKdV), and the KdV eigenfunction (KdV eig.). In addition,  in the commutative case \cite{BS1, apnum} also the Dym equation is included in  the 	B\"acklund chart.  On the other hand, in the non commutative one, as pointed out in \cite{EECT2019}, there are two different modified Korteweg-de Vries equations which are connected to each other via two further nonlinear evolution equations: both of them reduce to  the KdV eigenfunction when the non commutativity condition is removed.

\medskip
{{  Section \ref{background} is devoted to briefly recall the definition of B\"acklund transformation together with some remarkable properties. 
In the subsequent Section \ref{ab}   the   B\"acklund charts connecting Abelian
 KdV-type equations is provided. In particular, the B\"acklund chart in
  \cite{BS1, apnum} is recalled.
 The subsequent Section \ref{new-sol}  the non trivial  invariance exhibited by the KdV eigenfunction equation is used to construct explicit solution it admits. These solutions are independent on time, but show an explicit dependence on the space variable $x$. The extension of the  B\"acklund chart, induced by the M\"obius invariance enjoyed by the KdV singularity equation,  is given in Section \ref{extended}. 
  The Section \ref{nonab} reconsiders the non-Abelian B\"acklund chart, constructed in \cite{Carillo:Schiebold:JMP2009,SIGMA2016, JMP2018}. The last Section briefly discusses further remarkable results B\"acklund transformations allow to achieve. Matrix solutions, generalisation of results to hierarchies of nonlinear evolution equations are mentioned.  }}

\section{B\"acklund transformation \& B\"acklund charts}\label{background}
This Section is devoted to a brief review on B\"acklund transformations aiming to provide the definitions needed in the following
restriction the attention on the definitions used in the following 
%(see for instance \cite{} for details). 
The general definition of B\"acklund transformation in  implicit  form, according to \cite{RogersAmes} and references therein, 
given 
two non linear evolution equations of the type
\begin{equation}\label{1}
u_t = K ( u ) ~~, ~~  v_t =  G(v)
\end{equation}
when $\beta\! \in\!{\mathbb R}\setminus\{0\}$ denotes the B\"acklund parameter, then the B\"acklund Relations read
$$\left\{\begin{array}{l}
\displaystyle{{\partial v} \over{\partial x'}}=:{{\mathbb{B}'_1}(u,u_x;v;\beta),~~~~,~~~~ x' =  x'(x,t) } \\ \\
\displaystyle{{\partial v }\over{\partial t'}}=:{{\mathbb{B}'_2}(u,u_t;v;\beta)~~~~~~,~~~~  t'= t'(x,t)}
\end{array}\right.
$$
Then, compatibility conditions ${\displaystyle {\partial {\mathbb{B}^{\prime}_1} \over {\partial t^{\prime}}} =
   {{\partial {\mathbb{B}^{\prime}_2} }\over {\partial x^{\prime}}}}$ followed by elimination of $x',t',v$  from the latter 
 give  ${u_t = K (u)}$. On the other hand, when we write the B\"acklund Relations in terms of $(x, t)$, compatibility condition combined with elimination of $x,t,v$ produce $v_t =  G(v)$.
Throughout,  the following definition, see \cite{FokasFuchssteiner:1981}, is adopted.

\medskip\noindent
{\bf{ Definition}} 
{\it Given two evolution equations, 
{\begin{eqnarray*}
 u_t &= K ( u ),~ K : M_1 \rightarrow TM_1,~{{ u\!:(x,t)\!\!\in{\mathbb R}^n\!\times\!{\mathbb R}\!\to\! u (x,t)\! \in\!{\mathbb R}^m \! \! \subset M_1}}\\
 v_t &= G (v),~G : M_2  \rightarrow TM_2,~v\!:(x,t)\!\!\in{\mathbb R}^n\!\times{\mathbb R}\!\to\! v (x,t)\! \in\!{\mathbb R}^m \! \! \subset M_2
\end{eqnarray*} }
then $\hbox{B (u , v) = 0}$  
represents a B\"acklund transformation between them 
 whenever  given two solutions of such  equations, say, respectively, $u(x,t)$ and $v(x,t)$  such that 
\begin{equation}
B (u(x,t), v(x,t)) \vert_{ t=0 } = 0 
\end{equation}
it follows that,
\begin{equation}
B (u(x,t),v(x,t) )\vert_{t=\bar t} = 0,     ~~\forall \bar t >0 ~, ~~~\forall x\in{\mathbb R}.
\end{equation}
}

As usual choice when soliton solutions are considered, it is  assumed 
$M:= M_1\equiv M_2$ and, in addition, the generic 
{\it fiber} $T_uM$, at $u\in M$, is identified with $M$ itself\footnote{It is generally assumed that $M$ is the space of functions $u(x,t)$ which, 
for each fixed $t$, belong to the Schwartz space $S$ of {\it rapidly decreasing functions} on 
${\mathbb R}^n$,    i.e.
$S({\mathbb R}^n):=\{ f\in C^\infty({\mathbb R}^n) : \vert\!\vert f \vert\!
\vert_{\alpha,\beta} < \infty, \forall \alpha,\beta\in {\mathbb N}_0^n\}$, where 
$\vert\!\vert f \vert\!\vert_{\alpha,\beta}:= sup_{x\in{{\mathbb R}}^n} \left\vert x^\alpha D^\beta f(x)
\right\vert $, and  $D^\beta:=\partial^\beta /{\partial x}^\beta$; throughout this article $n=1$.}.  

\noindent As a consequence, solutions of such two equations are linked via the B\"acklund transformation 
which establishes a correspondence between them: it can  graphically represented 
as can be depicted by the following fugure %

\begin{eqnarray}\label{BC1}
\boxed{u_t = K (u)} \,{\buildrel B \over {\textendash\textendash}}\, \boxed{ v_t = G (v) }
\end{eqnarray}

\medskip\noindent
which shows that the B\"acklund transformation relates the two nonlinear evolution equations.  %

\section{Abelian B\"acklund chart}\label{ab}
The net of links connecting the many different nonlinear evolution equations can be summarised  in a B\"acklund chart; here the latest \cite{EECT2019} is reported. Indeed, the construction  is directly related to results in\cite{Fuchssteiner:Carillo:1989a, apnum} further developments, in the case of nc nonlinear evolution equations are comprised \cite{Carillo:Schiebold:JMP2009, Carillo:Schiebold:JMP2011, SIGMA2016, JMP2018} while a comparison between the two different cases is studied in \cite{EECT2019}.
The links among the various KdV-type equations  are summarised in the following  B\"acklund chart: 

 \begin{figure}[h]
\begin{gather*}
\mbox{$\small
\boxed{\!\text{KdV}(u)\!}\, {\buildrel (a)
\over{\text{\textendash\textendash}}} \, \boxed{\!\text{mKdV}(v)\! } \, {\buildrel (b) \over{\text{\textendash\textendash}}} \, \boxed{\!\text{KdV eig.}(w)\! } \, {\buildrel (c) \over{\text{\textendash\textendash}}} \, \boxed{\!\text{KdV~sing.}(\varphi)\!}
 {\buildrel (d) \over{\text{\textendash\textendash}}}\, \boxed{\!\text{int. sol KdV}(s) \!} \,
{\buildrel (e) \over{\text{\textendash\textendash}}} \, \boxed{\!\text{Dym}(\rho)\!}\,$}\label{BC1*}
\end{gather*}
\caption{KdV-type equtions B\"acklund chart}
\label{fig KdV-chart}
\end{figure}
\noindent
where, in turn,  the linked nonlinear evolution equations are: 
\begin{alignat*}{3} 
& u_t = u_{xxx} + 6 uu_x \qquad && \text{(KdV)}, & \\
& v_t = v_{xxx} - 6 v^2 v_x \qquad && \text{(mKdV)},& \\
& w_t = w_{xxx} - 3 {{w_x w_{xx}}\over w}\qquad && \text{(KdV eig.)},& \\
& \varphi_t = \varphi_x \{ \varphi ; x\} , \quad \text{where} \ \ \{ \varphi ; x \} :=
 \left( { \varphi_{xx} \over \varphi_x} \right)_x -
{1 \over 2 }\left({ \varphi_{xx} \over \varphi_x} \right)^2 \qquad && \text{(KdV~sing.)}, & \\
& s^2 s_t = s^2 s_{xxx} - 3 s s_x s_{xx}+ {3 \over 2 }{s_x}^3 \qquad && \text{(int.\ sol~KdV)}, & \\
& \rho_t = \rho^{3} \rho_{\xi \xi \xi} \qquad && \text{(Dym)}.&
\end{alignat*}

\medskip\noindent
that is, in  order, the Korteweg-de Vries (KdV), the modified Korteweg-de Vries (mKdV), and the KdV eigenfunction (KdV eig.), 
the Korteweg deVries interacting soliton (int.sol.KdV), the Korteweg deVries singuarity manifold (KdV sing.) and the Dym equations.
\noindent
Respectively, in the B\"acklund chart,  $(a), (b), (c), (d), (e)$ denote the following B\"acklund transformations
 \begin{gather}\label{links}
(a) \ \ u + v_x + v^2 =0 , \qquad \qquad \qquad (b) \ \ v -{{ w_{x} \over w} } = 0,\\
(c) \  w^2 -{  \varphi_x}  = 0,\!\qquad \qquad  \qquad\qquad (d) \ \ s - \varphi_x =0, 
\end{gather}
and 
\begin{equation}
(e) ~\ {\bar x} : = D^{-1} s (x), ~\rho(\bar x) :=  s(x), ~~~~~\text{where}~~~ ~D^{-1}:= \int_{-\infty}^x d\xi , \label{rec}
\end{equation}
so that  $\bar x= \bar x(s,x)$ and, hence, $ \rho(\bar x) :=  \rho(\bar x(s,x))$. The transformation  (e) is termed {\it reciprocal transformation} since it interchanges the role of the dependent and independent variables\footnote{see, for instance,  \cite{RogersShadwick} where  reciprocal transformations are defined and applications are provided. The transformation $(e)$ is 
analysed  in \cite{BS1, Fuchssteiner:Carillo:1989a} where it is shown to
represent a B\"acklund transformation between the extended manifold consisting of the both the dependent and the independent variables.}.

\section{The KdV eigenfunction equation: invariance properties and solutions' construction}\label{new-sol}

 The  {\it KdV eigenfunction} equation, for sake of brevity denoted as KdV eig., is included in a wide study by
 Konopelchenko in \cite{boris90} where, among many other ones,  it is proved to be integrable 
via  the {\it inverse spectral transform} (IST) method. Indeed, this equation was firstly 
derived  in a founding article of the IST method \cite{MGK} and also \cite{russi2}, later further investigated in  \cite{boris90, russi} wherein a  wide variety of nonlinear evolution equations is studied. Nevertheless, 
the KdV eigenfunction  equation does not appear in subsequent  classification studies of integrable nonlinear 
evolution equations, such as 
\cite{Calogero1985, Olver:Sokolov, Wang, Mikhailov-et-al, Mikhailov-et-al2} until very recently when, in \cite{Faruk1},  linearizable 
 nonlinear evolution equations are classified.
The KdV eigenfunction equation is a third order nonlinear equation of KdV-type since  it
is connected via B\"acklund  transformations with the  
Korteweg deVries (KdV),  the modified Korteweg deVries (mKdV), 
 the {\it Korteweg deVries interacting soliton} (int.sol.KdV) \cite{Fuchssteiner1987} and the 
 {\it Korteweg deVries singuarity manifold}  (KdV sing.),  introduced by Weiss in  \cite{Weiss} via the {\it Painlev\`e test} of integrability.
\subsection{Invariance}
 
This section is concerned only about  the  an invariance property  enjoyed by the  KdV eigenfunction equation. 
 It can be trivially checked to be {\it scaling invariant} since 
 on substitution of  $\alpha w, \forall \alpha \in \mathbb{C}$, to $w$  it remains unchanged.  
  In addition, according to \cite{apnum}, see prop. 4 therein, the following   proposition shows further nontrivial invariances.
\medskip

\noindent {\bf Proposition \ref{prop2b}\label{prop2b}} \\ \noindent  
{\it  The KdV eigenfunction equation $\displaystyle{w_t = w_{xxx} - 3 {{w_x w_{xx}}\over w}}$ is  invariant under the transformation 
\begin{equation}
\text{\rm I}: ~~~ \hat w^2 ={{ad- bc}\over{(c D^{-1}( w^2) +d)^2}}w^2,\quad a,b,c,d\in \mathbb{C}   
~ \text{s.t.} ~ad-bc\neq 0,  
\end{equation}
}
where
\begin{equation*}
D^{-1}:=\int_{-\infty}^x d\xi 
\end{equation*}
is well defined since so called {\it soliton solutions}  are looked for in the   Schwartz space $S({\mathbb R}^n)$ \footnote{see footnote on page 2.}.

The proof, according to \cite{apnum}, is based on the    invariance under the M\"obius group of 
transformations 
\begin{equation}
\text{M}:~~ \hat\varphi={{a\varphi+b}\over{c\varphi+d}},\qquad a,b,c,d\in \mathbb{C} \qquad \text{such that} \quad ad-bc\neq 0.
\end{equation}
of the KdV singularity manifold equation 
\begin{equation}
\varphi_t = \varphi_x \{ \varphi ; x\} , \quad \text{where} \ \ \{ \varphi ; x \} :=
 \left( { \varphi_{xx} \over \varphi_x} \right)_x -
{1 \over 2 }\left({ \varphi_{xx} \over \varphi_x} \right)^2.
\end{equation}
Combination of such an invariance with the connection between the KdV eigenfunction and
 the KdV singularity manifold 
equation allows to prove  the proposition. 
Indeed, let
\begin{equation}
\text{M}: \hat \varphi={{a\varphi+b}\over{c\varphi+d}}~~,~~ \forall a,b,c,d\in \mathbb{C} \vert ~ad-bc\ne 0
\end{equation}
 the following B\"acklund chart % \medskip
 \begin{figure}[h]
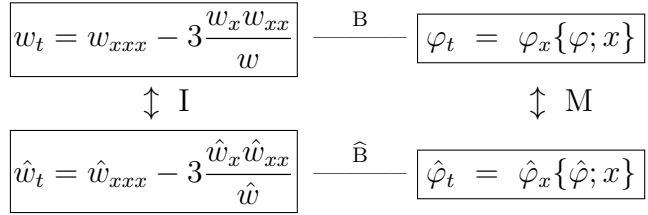

 \begin{eqnarray*}
\boxed{\!w_t = w_{xxx} - 3 {{w_x w_{xx}}\over w}\! } ~\, {\buildrel {\rm B}\over{\text{\textendash\textendash\textendash\textendash\textendash\textendash}}}~\boxed{\varphi_t \ =\  \varphi_x  \{ \varphi ; x\}}\\
\updownarrow~ \text{I} ~~~\qquad\qquad\qquad\qquad~~~~~\updownarrow~ \text{M} ~~~~~ \\
\boxed{\!\hat w_t = \hat w_{xxx} - 3 {{\hat w_x \hat w_{xx}}\over \hat w}\! } \,~ {\buildrel\widehat {\text{\rm B}}\over{\text{\textendash\textendash\textendash\textendash\textendash\textendash}}}~\boxed{\hat\varphi_t \ =\  \hat \varphi_x  \{ \hat \varphi ; x\}}\\
\end{eqnarray*}
\caption{Induced invariance  B\"acklund chart.}
\label{fig KdV-eig-chart}
\end{figure}
where the B\"acklund transformations ${\rm B}$ and $\widehat{\rm B}$ are, respectively:
\begin{equation*}
\text{\rm B}: ~~~~\displaystyle{w^2 -  { \varphi_x}=0~ }~~~~\text{and}~~~~~~~~~~\widehat {\text{\rm B}}: ~~~~ \displaystyle{\hat w^2 -  {\hat \varphi_x}=0~.}
\end{equation*}
The invariance I follows via combination of the M\"obius   transformation M with the two B\"acklund transformations ${\rm B}$ and $\widehat{\rm B}$.
An application of the invariance ${\rm I}$ is indicates how to construct solutions of the KdV eigenfunction equation.

\subsection{Explicit Solutions: an example}
In this subsection an example of solutions admitted be the KdV eigenfunction equation is constructed on the basis of the invariance in the previous subsection. Indeed, it is easily checked that $w(x,t)=k_1, \forall k_1\in {\mathbb R}$ represents a solution of the KdV eigenfunction equation
\begin{equation}
w_t = w_{xxx} - 3 {{w_x w_{xx}}\over w}~.
\end{equation}\label{w}
When, in the M\"obius group the parameters are set to be
\begin{equation}
a=d=0, b=1, c=-1
\end{equation}
the invariance $I$ indicates that also  
\begin{equation}
\hat w(x,t)= (k_1^2 x+k_2)^{-1}~,  ~~ \forall k_2\in{\mathbb R}
\end{equation}
represent  solutions of the KdV eigenfunction equation.

Further solutions  can be obtained in the same way. Remarkably, also in the nc case solutions can be constructed.

\section{Extension of the Abelian B\"acklund chart}\label{extended}
 
Note that   the whole  B\"acklund chart in Fig. \ref{fig KdV-chart} can be extended, as  indicated in the following figure.

 \begin{figure}[h]
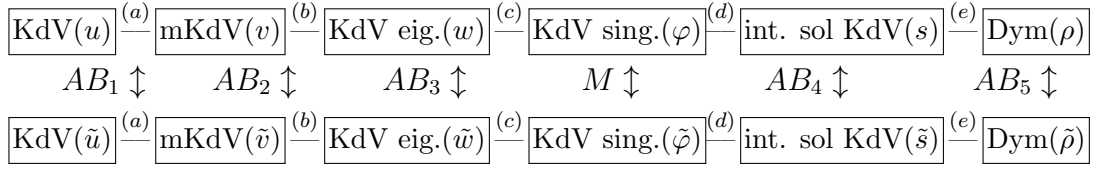


\begin{eqnarray*}
\mbox{\small $ \boxed{\!\text{KdV}(u)\!}\, {\buildrel (a)
\over{\text{\textendash\textendash}}} \, \boxed{\!\text{mKdV}(v)\! } \, {\buildrel (b) \over{\text{\textendash\textendash}}} \, \boxed{\!\text{KdV eig.}(w)\! } \, {\buildrel (c) \over{\text{\textendash\textendash}}} \, \boxed{\!\text{KdV~sing.}(\varphi)\!}
 {\buildrel (d) \over{\text{\textendash\textendash}}}\, \boxed{\!\text{int. sol KdV}(s) \!} \,
{\buildrel (e) \over{\text{\textendash\textendash}}} \, \boxed{\!\text{Dym}(\rho)\!}\,$}
 \\ \small 
AB_1\updownarrow~ ~~~ ~~AB_2 \updownarrow~~~~~~~~AB_3 \updownarrow~~~~~~~~~~~M \updownarrow~~~~~~~~~~~~AB_4 \updownarrow~~~~~~~~~~~~AB_5 \updownarrow~~~ \\
\mbox{\small $\boxed{\!\text{KdV}(\tilde u)\!}\, {\buildrel (a)
\over{\text{\textendash\textendash}}} \, \boxed{\!\text{mKdV}(\tilde v)\! } \, {\buildrel (b) \over{\text{\textendash\textendash}}} \, \boxed{\!\text{KdV eig.}(\tilde w)\! } \, {\buildrel (c) \over{\text{\textendash\textendash}}} \, \boxed{\!\text{KdV~sing.}(\tilde \varphi)\!}
 {\buildrel (d) \over{\text{\textendash\textendash}}}\, \boxed{\!\text{int. sol KdV}(\tilde s) \!} \,
{\buildrel (e) \over{\text{\textendash\textendash}}} \, \boxed{\!\text{Dym}(\tilde \rho)\!}\,$}
\end{eqnarray*}
\caption{Abelian KdV-type hierarchies  B\"acklund chart: induced invariances.}
\label{fig KdV-ext-chart}

\end{figure}
\noindent
that is, since the KdV singularity manifold equation (KdV \-sing.) is invariant under the M\"obius group of transformations, all the the  auto-B\"acklund transformations $AB_k, k= 1\dots 5$ follow. Note that  AB$_1$ and AB$_2$ are, respectively, the well known KdV and the mKdV auto-B\"acklund transformations  \cite{Miura, CalogeroDegasperis, Fuchssteiner:Carillo:1989a}. The invariance of the KdV eigenfunction equation is  I$\equiv AB_3$ \cite{apnum} and, according to  \cite{Fuchssteiner:Carillo:1989a}, auto-B\"acklund transformations of the 
the int. sol. KdV and Dym equations are also obtained, denoted as  AB$_4$, and  AB$_5$. 
  
\section{Non-Abelian B\"acklund chart}\label{nonab}
In this section the attention is focussed on non-Abelian equations. Specifically, according to \cite{Marchenko, AdenCarl, CarlSchiebold} 
 nonlinear evolution equations in which the unknown is an operator on a~Banach space are studied. These, for short, are termed {\it operator equations}. Crucial to the present study, both in the Abelian as well as in the non-Abelian setting,   is that the algebraic   properties of interest nonlinear evolution equations enjoy are preserved under B\"acklund transformations. To stress the distinction between scalar and operator unknown functions, they are, respectively, denoted via lower and upper-case letters. 
Taking into account the results in 
\cite{AF, Carillo:Schiebold:JMP2009, SIGMA2016, JMP2018}, a B\"acklund chart which connects operator KdV-type equations, can be constructed: it  is depicted in the following   Fig. \ref{fig nc KdVchart}.
 
 \begin{figure}[h]

\unitlength.8cm
\begin{picture}(15,6.75)
   
% nc
   \put(0.5,5.5){\framebox(2,0.5){\shortstack{\footnotesize KdV$(U)$}}}
   \put(4,5.5){\framebox(2.5,0.5){\shortstack{\footnotesize mKdV$(V)$}}}
   
   \put(1.3,3){\framebox(2.5,1){\shortstack{\footnotesize \\ \scriptsize meta-mKdV$(Q)$}}}
    \put(6,3){\framebox(2.5,1){\shortstack{\scriptsize mirror \\ \scriptsize meta-mKdV$(\widetilde Q)$}}} 
 
   \put(4,0.5){\framebox(2.5,1){\shortstack{\footnotesize alternative \\ \footnotesize mKdV$(\widetilde V)$}}}
   \put(8,0.75){\framebox(2.5,0.5){\shortstack{\footnotesize Int So KdV $(S)$}}}
   \put(12,0.75){\framebox(2.5,0.5){\shortstack{\footnotesize KdV Sing $(\phi)$}}}

   \put(3.75,5.75){\vector(-1,0){1}}   \put(3.1,6){\footnotesize $(a)$}
  
   \put(3.5,4.25){\vector(1,1){1}}		 \put(3.4,4.75){\footnotesize $\hskip-4em  V= Q_xQ^{-1}$}
   \put(3.5,2.75){\vector(1,-1){1}}	 \put(3.4,2){\footnotesize $\hskip-5em \widetilde V= - \  \widetilde Q_x\widetilde Q^{-1}$}
   \put(6.5,4.25){\vector(-1,1){1}} 	 \put(6.2,4.75){\footnotesize $  V= - \ \widetilde Q^{-1}\widetilde Q_x$}
   \put(6.5,2.75){\vector(-1,-1){1}}    	 \put(6.2,2){\footnotesize $\widetilde V= - \  \widetilde Q_x\widetilde Q^{-1}$}
   
   \put(7.75,1){\vector(-1,0){1}}        \put(7.1,1.2){\footnotesize $(b)$}
   \put(11.75,1){\vector(-1,0){1}}      \put(11.1,1.2){\footnotesize $(d)$}

\end{picture}
\caption{KdV-type  equations  B\"acklund chart: the non-Abelian case.}
\label{fig nc KdVchart}

\end{figure}
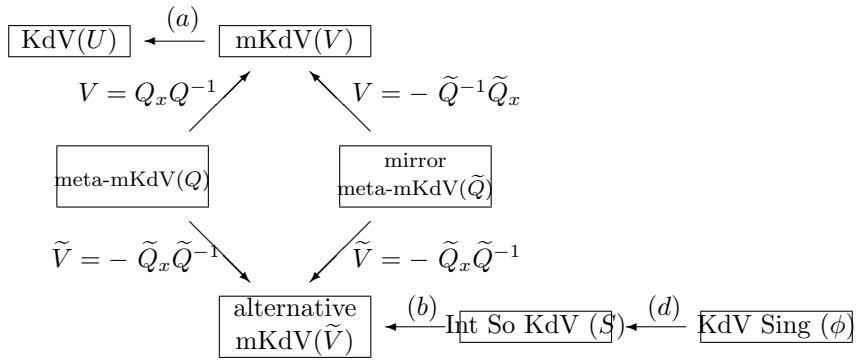

\noindent In Fig. \ref{fig nc KdVchart}, the third order nonlinear operator evolution equations where, as pointed out, all unknowns are denoted via capital case letters with the only exception of the KdV sing. equation. The KdV-type 
operator equations  are, in turn,
\begin{alignat*}{3} 
& U_t = U_{xxx} + 3 \{ U, U_x \} \qquad && \text{(KdV)}, & \\
&  V_t = V_{xxx} - 3 \{ V^2, V_x \} \qquad && \text{(mKdV)},& \\
&   Q_t = Q_{xxx} -3Q_{xx}Q^{-1}Q_{x} && \text{(meta-mKdV)},& \\
&   \widetilde Q_t = \widetilde Q_{xxx} -3\widetilde Q_{x} \widetilde Q^{-1}\widetilde Q_{xx} && \text{(mirror meta-mKdV)},& \\
&   \widetilde V_t = \widetilde V_{xxx} + 3 [ \widetilde V, \widetilde V_{xx} ] -6 \widetilde V \widetilde V_x \widetilde V\qquad && \text{(amKdV)},& \\
&  \phi_t = \phi_{x} \{ \phi; x \}  , \quad \text{where} \ \    \{ \phi ; x \} = \big( \phi_x^{-1} \phi_{xx} \big)_x - \frac{1}{2} \big( \phi_x^{-1} \phi_{xx} \big)^2 \qquad && \text{(KdV~sing.)}, & \\
& S_t = S_{xxx} - \frac{3}{2} \big( S_xS^{-1}S_x \big)_x  \qquad && \text{(int.\ sol~KdV)}. &
\end{alignat*}
where the B\"acklund transformations $(a), (b), (c)$  linking the KdV with the mKdV, the amKdV  with the  int.sol~KdV and the latter  with the   KdV~sing. are the following, ones, the  non-commutative counterparts of those in the commutative B\"acklund chart, see  Fig. \ref{fig KdV-chart} and Fig.  \ref{fig KdV-ext-chart}. 
\begin{alignat*}{3} 
&   U = -\  (V^2+V_x) \hfill &&  {(a)} & \\
&    \widetilde V = \frac{1}{2} S^{-1} S_x \qquad\qquad\qquad\qquad\qquad\qquad\qquad &&  {(b)}& \\
&      S = \phi_x ~.&&  {(c)}~.
\end{alignat*}
Notably, both the two equation named mirror meta-mKdV and meta-mKdV \cite{JMP2018} coincide with the 
KdV eigenfunction equation when the nc condition is removed. Similarly, also the mKdV and amKdV equations 
when commutativity is assumed reduce to the usual mKdV equation and, therefore the {\it box} in Fig. \ref{fig nc KdVchart} collapses to a line and, therefore the B\"acklund chart in Fig. \ref{fig KdV-chart} can be recovered. The following   Fig. \ref{meta-box} is esplicative.
\begin{center}
{\includegraphics[height=1.5in,width=3.5in]{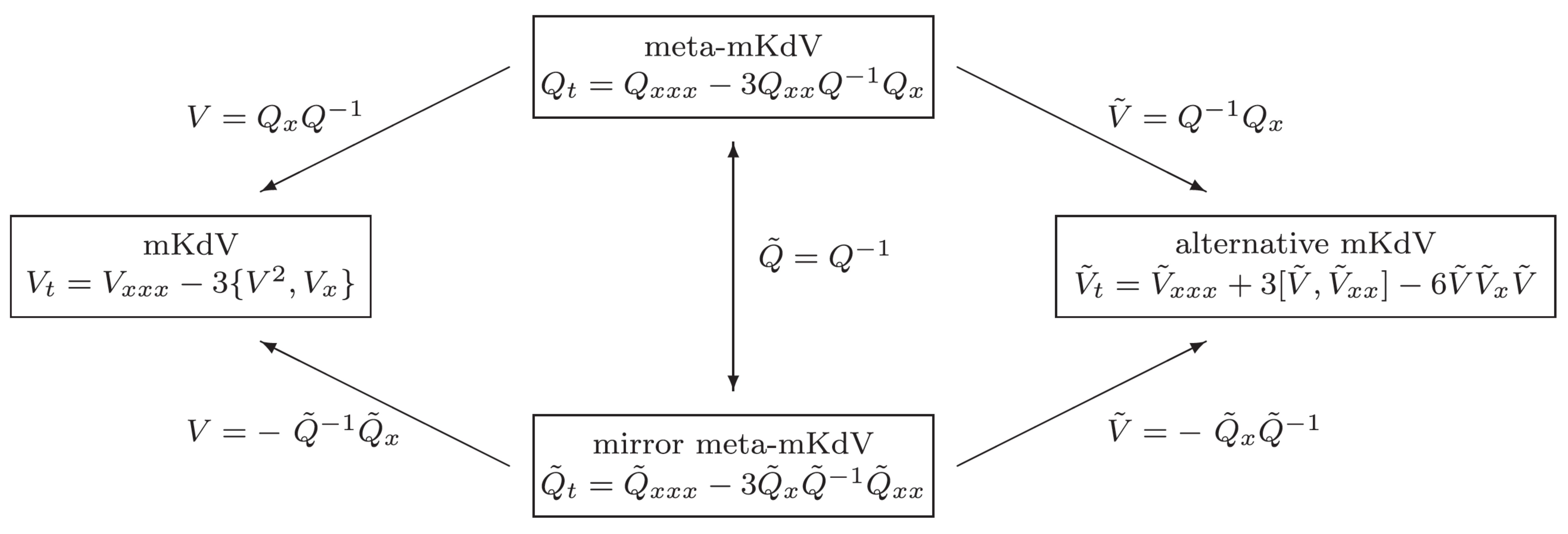}}\label{meta-box}
%\caption{Fig. 5 mKdV  B\"acklund chart: a closer look.}\label{meta-box}
  \end{center}

\noindent
where note that, in the commutative case, the composition of the transformations    
$V= Q_xQ^{-1}$ and     $  \widetilde V= - \ Q^{-1} Q_x$ reduce to $V \to -V$, 
trivial sign invariance admitted by the  mKdV equation;  correspondingly, the two 
forms of modified KdV equations (amKdV and mKdV) coincide with the (Abelian)
 mKdV equation. 
\section{Conclusions and perspectives}\label{perspectives}
This closing section aims to give  a brief overview on two different lines of results concerning further developments  that is, on one side the determination of explicit  solutions  and on te other one, the extension  of the   B\"acklund 
chart from the considered nonlinear evolution equations to corresponding hierarchies.  Further perspectives are finally mentioned.

\subsection{Matrix solutions of soliton equations}\label{matrix}
This subsection is devoted to the special case when the operator is finite dimensional so that it admits a matrix representation. Thus, the aim is to emphasise the importance of B\"acklund transformations also when  solutions admitted by non-Abelian soliton equations are looked for.
 Solutions admitted by the matrix  equations are a subject of interest in the literature. The study presented, based on previous results \cite{Carillo:Schiebold:JMP2009, Carillo:Schiebold:JMP2011} further developed in \cite{NODYCON, NODYCON2}, 
 take into account multisoliton solutions of the matrix KdV equation obtained by Goncharenko \cite{Goncharenko},  via a generalisation of the Inverse Scattering Method.  According to \cite{Carillo:Schiebold:JMP2011}, and in particular Theorem 3 therein, generalises 
 Goncharenko's multisoliton solutions which follow as special ones. Solutions of matrix mKdV equation are discussed and obtained in \cite{NODYCON}, where some $2 \times 2$ and some $3 \times 3$ examples are provided; in \cite{NODYCON2}  the solution formula, which in\cite{Carillo:Schiebold:JMP2011}  is obtained in the general operator case, is discussed referring to the case of a $d \times d, d\in\mathbb{N}$ and further solutions are produced 
to give an idea of the results in this direction and currently under investigation \cite{preprint}. 
Further matrix  solutions are obtained  in \cite{Chen, Goncharenko, LRB, Schiebold2009, Schiebold2018, Hamanaka1, Sakh}.

\subsection{Generalisations: hierarchies \& further perspectives}\label{hierarchies}
Finally, some further observations deserve attention. First of all, one of the properties of B\"acklund transformations crucial in the present research project, which involves not only the two authors, but also further collaborators, is the notion of {\it recursion operator}.
Indeed the existence of a hereditary recursion operator admitted by a nonlinear evolution equation is a remarkable algebraic property \cite{Fuchssteiner1979}. Such a property, on one side, allows to construct a whole {\it hierarchy} of nonlinear evolution equations associated, for instance,  to the KdV equation. On another side, the algebraic properties which characterise  a hereditary recursion operator are preserved under B\"acklund transformations. Hence, since the KdV equation admits  a hereditary recursion operator, the B\"acklund chart in Fig.\ref{fig KdV-chart} indicates the way to construct  the recursion operators of all the nonlinear evolution equations it connects. In addition, such a B\"acklund chart   can be naturally extended to the whole hierarchies of all the nonlinear evolution equations therein \cite{Fuchssteiner:Carillo:1989a}, in the Abelian case.The corresponding non-Abelian B\"acklund chart is constructed in \cite{Carillo:Schiebold:JMP2009, Carillo:Schiebold:JMP2011} and further extended in \cite{SIGMA2016, JMP2018}.  
Also the case of non-Abelian Burgers equation \cite{Ku,  GKT, Hamanaka2, Carillo:Schiebold:JNMP2012, MATCOM2017} shows a richer structure with respect to the corresponding Abelian one. 

Furthermore, a $2+1$- dimensional \cite{walsan1}   B\"acklund chart which links the Kadomtsev-Petviashvili (KP), the modified Kadomtsev-Petviashvili (mKP) and other $2+1$- dimensional soliton equations, such as the KP singularity manifold equation and the  
$2+1$- dimensional version of the Dym equation. The link, obtained by Rogers  \cite{Rogers:1987}, between the KP and the $2+1$-dimensional Dym equation indicates the way to the construction of solutions of the Dym equation in $2+1$- dimensions.   Notably, as shown in \cite{walsan1}, the BC in Fig. \ref{fig KdV-ext-chart}  follows to represent a
{\it constrained} version of the KP  B\"acklund chart   \cite{walsan2}.  In addition, as shown in  \cite{Fuchssteiner:Carillo:1989a}, the Hamiltonian and bi-Hamiltonian structure admitted by the KdV equation, since these properties are preserved under B\"acklund transformations \cite{Fuchssteiner1979}, all    the nonlinear evolution KdV-type equations in the B\"acklund chart,  in Fig. \ref{fig KdV-ext-chart}, are all proved to admit a bi-Hamiltonian structure 
\cite{Magri,[12], Fuchssteiner:Carillo:1990a, Benno-Walter}. As discussed in \cite{ActaAM2012}, a B\"acklund chart connects the Caudrey-Dodd-Gibbon-Sawata-Kotera and Kaup-Kupershmidt hierarchies  \cite{CDG, SK, Kawa}.  All the involved equations are 5th order  nonlinear evolution equations; notably,  the B\"acklund chart linking them all shows an impressive resemblance to the one connecting KdV-type equations. Again, such B\"acklund chart can be extended to the corresponding whole hierarchies \cite{Rogers:Carillo:1987b, BS1}. The study on $2+1$-dimensional non-Abelian equations seems of interest.

\index{running head}

\section*{Acknowledgments}
The financial support of G.N.F.M.-I.N.d.A.M.,  I.N.F.N. and \textsc{Sapienza}  University of Rome, Italy are gratefully acknowledged. C. Schiebold thanks Dipartimento Scienze di Base e Applicate
    per l'Ingegneria and \textsc{Sapienza}  University of Rome for the kind hospitality. 
%%%%%%%%%%%%%%%%%%%%%%%%%%%%%%%%%%%%%%

%\end{verbatim}

\begin{thebibliography}{9}
\bibitem{Ablowitz} M.J. Ablowitz, D.J. Kaup, A.C. Newell and H. Segur,  The inverse scattering transform - Fourier analysis for nonlinear problems, {\it{Studies in Applied Mathematics}}, Vol. 53, 249-315 (1974). 

\bibitem{AdenCarl} %cited
Aden H., Carl B., On realizations of solutions of the {K}d{V} equation by determinants on operator ideals, \href{http://dx.doi.org/10.1063/1.531482}{\textit{J.~Math. Phys.}} \textbf{37} (1996),
 1833--1857.
 
\bibitem{AF} %NO cited
Athorne C., Fordy A., Generalised {K}d{V} and {MK}d{V} equations associated with symmetric spaces, \href{http://dx.doi.org/10.1088/0305-4470/20/6/021}{\textit{J.~Phys.~A: Math. Gen.}} \textbf{20} (1987),
 1377--1386.

\bibitem{Faruk1}        %ok
P. Basarab-Horwath and , and F. G\"ung\"or,
 Linearizability for third order evolution equations,
  {\it{   J. Math. Phys.}} {\bf 58} 
081507 (2017); \\ \doi{ 10.1063/1.4997558 } 

\bibitem{Calogero1985} %ok
F. Calogero and A. Degasperis, 
A modified modified Korteweg-de Vries equation, 
{\it{Inverse Problems}} {\bf 1}, (1985), 57-66.

\bibitem{CalogeroDeLillo} %ok
F.~Calogero and S. De Lillo, 
The Burgers equation on the semiline with general boundary conditions at the origin, 
{\it{J. Math. Phys.}} {\bf 32}, 1 (1991),   99-105. 

\bibitem{CalogeroDegasperis} %ok
F.~Calogero and  A.~Degasperis,
Spectral Transform and Solitons I,
{\it{Studies in Mathematics and its Application}}, Vol. 13,
North Holland, Amsterdam, 1980. 

\bibitem{ActaAM2012}  S~Carillo, %ok      
{Nonlinear Evolution Equations: B\"acklund Transformations and B\"acklund charts},\ 
{\it{Acta Applicandae Math.}} {\bf 122},   1 (2012),  93-106. %(9729)
\doi{ 10.1007/s10440-012-9729-8}

 \bibitem{apnum} %ok
 {  S. Carillo},  {\it  KdV-type equations linked via B\"acklund transformations:  remarks and perspectives} 
Applied Numerical Mathematics,    {\bf 141},  81--90, (2019).

\bibitem{BS1} %ok
     S.~Carillo and B.~Fuchssteiner, 
        {The abundant symmetry structure of hierarchies of nonlinear equations obtained by reciprocal links,}
   {\it{   J. Math. Phys.}} {\bf 30}   (1989), {1606-1613}.
   \doi{10.1016/j.apnum.2018.10.002}
   
%            \bibitem {BSKonop}  % removed: not cited
%            S.  Carillo, B. Fuchssteiner  and G.B.
%        Konopelchenko   \ {\sl  The action-angle
%        transformation for interacting solitons and the
%        dynamics of eigenfunctions for soliton
%        equations}, \ Rend. Mat., Serie VII, {\bf II},
%        (1991),  201-226.
%    \bibitem{SIMAI2008}   %ok  not cited
%    S.~Carillo and  C. Schiebold,  { A non-commutative operator-hierarchy of Burgers equations and B\"acklund transformations}, in {\it{Applied and Industrial Mathematics in Italy III
%    %: Selected Contributions from the 9th SIMAI Conference
%    }},     E. De Bernardis, R.
%    Spigler, V. Valente Ed.s, SERIES ON ADV. MATH.
%    APPL. SCIENCES, Vol.82,   World Scientific Pubbl., 
%    Singapore, 2010, 175 -185.

 \bibitem{Carillo:Schiebold:JMP2009} %ok
          S.~Carillo and C.~Schiebold,
          {Non-commutative KdV and mKdV hierarchies via recursion methods,}
       {\it{   J. Math. Phys.}} {\bf 50} (2009), 073510.

   \bibitem{Carillo:Schiebold:JMP2011} %ok
          S.~Carillo and C.~Schiebold,
          {Matrix Korteweg-de Vries and modified Korteweg-de Vries hierarchies:
                        Non-commutative soliton solutions,}
          {\it{J. Math. Phys.}} {\bf 52}  (2011), 053507.
   
\bibitem{Carillo:Schiebold:JNMP2012}  %ok NO
 S. Carillo,   C. Schiebold, 
{ On the recursion operator for the non-commutative Burgers hierarchy},\    
{\it{ J. Nonlin. Math.  Phys.}}\!\!   {\bf 19}\!\! 1  (2012), 1250003, 11 pp.
\doi{ 10.1142/S1402925112500039}

\bibitem{SIGMA2016} %ok
S.~Carillo, M.~Lo Schiavo,  C.~Schiebold,
 {B\"acklund Transformations \!and\! Non Abelian Nonlinear Evolution Equations: a novel B\"acklund chart},
{\it{ SIGMA}} 12 (2016), 087, 17 pages. 
\doi{10.3842/SIGMA.2016.087} 

\bibitem{MATCOM2017} %ok NO
S.~Carillo, M.~Lo Schiavo,  C.~Schiebold,
 {Recursion Operators admitted by non-Abelian Burgers equations: Some Remarks},
 {\it Math. and Comp. in Simul.}, {\bf 147C} (2018),   40-51. 
\doi{10.1016/j.matcom.2017.02.001}

\bibitem{JMP2018} %ok
S.~Carillo, M.~Lo Schiavo, E. Porten,  C.~Schiebold,
 {  A novel noncommutative KdV-type equation, its recursion operator, and solitons}, 	
\  {\it J. Math. Phys.},  {\bf 59}  (3), (2018),   3053--3060.

\bibitem{EECT2019} 
	\newblock S. Carillo, M. Lo Schiavo, and C. Schiebold.
   	\newblock \emph{Abelian versus non-Abelian B\"acklund charts: Some remarks.}  
	\newblock Evolution Equations and Control Theory {\bf 8}, 43--55 (2019). 	
	
\bibitem{NODYCON} 
	\newblock S. Carillo, M. Lo Schiavo, and C. Schiebold.
	\newblock \emph{Matrix solitons solutions of the modified Korteweg-de Vries equation.}
	\newblock In: Nonlinear Dynamics of Structures, Systems and Devices, 
		edited by W. Lacarbonara, B. Balachandran, J. Ma, J. Tenreiro Machado, G. Stepan.
		(Springer, Cham, 2020), pp. 75--83
				
\bibitem{NODYCON2} 
	\newblock S. Carillo,  and C. Schiebold.
	\newblock \emph{Construction of soliton solutions of the matrix modified Korteweg-de Vries equation
.}
	\newblock In: Nonlinear Dynamics of Structures, Systems and Devices, 
		edited by W. Lacarbonara, et al.
		(Springer, Cham, 2021), in press.
				
\bibitem{preprint}	
	\newblock S. Carillo, M. Lo Schiavo, and C. Schiebold.
	\newblock \emph{$N$-soliton matrix mKdV solutions: a step towards their classification.}
	\newblock Preprint 2020.	

   \bibitem{CarlSchiebold} %ok
   B.~Carl,  C.~Schiebold,
      {Nonlinear equations in soliton physics and operator ideals},
   {\it{Nonlinearity}} {\bf 12} (1999), 333-364.
   
    \bibitem{CDG} P. J. Caudrey, R. K. Dodd,  J. D. Gibbon,
A New Hierarchy of Korteweg-de Vries Equations,
  {\it Proc. Roy. Soc. London},  {\bf A 351}  (1976),  407-422.
   
\bibitem{Chen}
	\newblock X. Chen, Y. Zhang, J. Liang, and R. Wang.
	\newblock \emph{The N-soliton solutions for the matrix modified Korteweg-de Vries equation
		via the Riemann-Hilbert approach.}
	\newblock Eur. Phys. J. Plus (2020), 135:574.

%    \bibitem{Cole:1951} %ok NO BURG
%    {J.D.~Cole},
%     {On a quasilinear parabolic equation occuring in aerodynamics},
%    {\it{Quart. App. Math.}} {\bf 92} (1951), {25-236}.
	
%\bibitem{Depireux} {\red\boxed{NON }}
%	D.A.~Depireux,  J.~Schiff,
%	{On UrKdV and UrKP,}	
%	{\it{Lett. Math. Phys.}} {\bf 33} (1995), 99-111.

\bibitem{FokasFuchssteiner:1981} %ok
      A.S.~Fokas,  B.~Fuchssteiner,
      {B\"acklund transformation for hereditary symmetries,}
      {\it{Nonlin. Anal., Theory Methods Appl.}} {\bf 5}, 4 (1981), 423-432.

   \bibitem{Fuchssteiner1979} %ok
          B.~Fuchssteiner,
          {Application of hereditary symmetries to nonlinear evolution equations,}
          {\it{Nonlin. Anal., Th. Meth. Appl. }}{\bf 3},   6 (1979), 849-862.
%   \bibitem{Fuchssteiner1987} {\red\boxed{NON }}
%          B.~Fuchssteiner,
%Solitons in interaction,
%  {\it{Progr. Theor. Phys.}}{\bf 78}    5 (1987), 1022-1050.     
%     
  \bibitem{[12]}  %ok
  B.~Fuchssteiner, 
  The Lie algebra structure of Degenerate Hamiltonian and Bi-Hamiltonian systems,
  {\it{Progr. Theor. Phys.}},        {\bf 68}  (1982),  1082-1104.
   \bibitem{Fuchssteiner1987} 
          B.~Fuchssteiner,
Solitons in interaction,
  {\it{Progr. Theor. Phys.}}{\bf 78}    5 (1987), 1022-1050.     
      
\bibitem{Fuchssteiner:Carillo:1989a} %OK
{B.~Fuchssteiner,  S.~Carillo},
{Soliton structure versus singularity analysis: Third order completely integrable nonlinear equations in 1+1 dimensions,}
{\it{{Phys.} A }}  {\bf 154} (1989),  {467-510}.

\bibitem {Fuchssteiner:Carillo:1990a} % ok  {\red\boxed{NON }}
{B.~Fuchssteiner,  S.~Carillo}, {The Action-Angle Transformation for Soliton Equations}, \
 {\it{Physica A}}, {\bf 166}   (1990), 651-676.
 \doi{10.1016/0378-4371(90)90078-7}
%      
%\bibitem {BS4}  % ok 
%B.~Fuchssteiner, T.~Schulze,  S.~Carillo,
% \ { Explicit solutions for the Harry Dym equation}, 
%\ {\it{J.  Phys. A}} {\bf 25}  (1992), 223-230.

   \bibitem{Benno-Walter}    %ok {\red\boxed{NON }}
      B.~Fuchssteiner,  W.~ Oevel, 
The bi-Hamiltonian structure of some nonlinear fifth- and seventh-order 
differential equations and recursion formulas for their symmetries and conserved covariants, 
{\it{J. Math. Phys}} {\bf 23}  3 (1982),  358-363.


\bibitem{Goncharenko}
   	\newblock V.M. Goncharenko.
	\newblock \emph{Multisoliton solutions of the matrix KdV equation.}
	\newblock Theor. Math. Phys. {\bf 126} (2001), 81--91.

%
%\bibitem{Guo:Carillo} %ok
%{B. Yu Guo and S. Carillo},
% Infiltration in soils with prescribed
%boundary concentration: a Burgers model, 
%{\it{ Acta Appl. Math. Sinica}} {\bf 6},  N.4, (1990),   pp. 365-369.
%
%\bibitem{Guo:Rogers}  % ok
%{B.Yu~Guo,  C.~Rogers, On the Harry Dym Equation and its Solution, 
%{\it{Science in China}} 32 (1989), 283-295.}
     
 \bibitem{Gu-book} %ok
    {C. Gu, H. Hu,  Z. Zhou},
  {Darboux transformations in integrable systems},
{Springer, Dordrecht}, {2005},      ISBN  {1-4020-3087-8}.

   \bibitem{GKS} %ok
       M.~G\"urses, A.~Karasu, and V.V.~Sokolov.
  {On construction of recursion operators from Lax representation},
     {\it{ J. Math. Phys.}} {\bf 40}, 6473--6490 (1999).

   \bibitem{GKT} %ok
    M.~G\"urses, A.~Karasu, and R.~Turhan.
{On non-commutative integrable Burgers equations},
 {\it{ J. Nonlinear Math. Phys.}} {\bf 17},  (2010) 1--6.
      
\bibitem{Hamanaka1} %OK
M. Hamanaka, Noncommutative solitons and quasideterminants, {\bf 89},  (2014) {\it{Phys. Scr.}} 038006.

\bibitem{Hamanaka2} %OK
M. Hamanaka  and K. Toda, Noncommutative Burgers equation, {\bf 36},  (2003) {\it{J. Phys. A: Math. Gen.}} 11981.
%\bibitem{Hopf:1950} %ok
%{E.~Hopf},
% {The partial differential equation $ u_{t} + u u_{x} = m
%uu_{xx}$,}
%{\it{Comm. Pure Appl. Math.}}
%{\bf 3}  (1950), {201-230}.
         
\bibitem{Kawa} %ok 
S. Kawamoto,
An exact Transformation from the Harry Dym Equation to the modified KdV Equation,
{\it{ J.Phys.Soc. Japan}} {\bf 54} (1985),2055-2056.

\bibitem{boris90} %OK
 B.G. Konopelchenko,
Soliton eigenfunction equations: the IST integrability and some properties,
{\it{Rev. Math. Phys.}} {\bf 2} 4, (1990),  399-440. 		
         
\bibitem{KK} %OK
   F.A.~Khalilov, E.Ya.~Khruslov.
   {Matrix generalisation of the modified Korteweg-de Vries equation.}
{\textit{Inv. Problems}} {\bf 6}, 193-204 (1990).

   \bibitem{Ku} %ok
 B.A.~Kupershmidt.
{On a group of automorphisms of the noncommutative Burgers hierarchy.}
      {\it{ J. Nonlinear Math. Phys.}} {\bf 12}, No. 4, 539-549 (2005).
%
%\bibitem{Leo} %ok
%M. Leo, R.A. Leo, G.  Soliani, L. Solombrino,
%On the isospectral-eigenvalue problem and the recursion operator of the Harry Dym equation. 
%{\it{Lett. Nuovo Cimento}} (2) {\bf 38}  2, (1983),  45-51.

   \bibitem{LRB} %ok
 D.~Levi, O.~Ragnisco, and M.~Bruschi.
 {Continuous and discrete matrix Burgers' hierarchies,}
     {\it{Il Nuovo Cimento}} {\bf 74B}, 33-51 (1983).

%   \bibitem{Lou} %ok
%          S.Y.~Lou,
%{Symmetries and similarity reductions of the Dym equation. }
%{\it Phys. Scripta}  {\bf 54} 5 (1996),   428-435.

 \bibitem{Magri}  %ok
 F.~Magri, A simple model of the integrable Hamiltonian equation,
 {\it{J.Math.Phys.}},  {\bf 19} (1978),   1156-1162.

\bibitem{Marchenko} %ok
Marchenko V.A., Nonlinear equations and operator algebras, %\href{http://dx.doi.org/10.1007/978-94-009-2887-9}
{\it{Mathematics and its Applications (Soviet Se\-ries)}}, Vol.~17, D.~Reidel Publishing Co., Dordrecht, 1988.

 \bibitem{russi} %OK
 A. G. Meshkov, V. V. Sokolov, Integrable evolution equations with a constant separant,  {\it{Ufimsk. Mat. Zh.}},  {\bf 4}, 3 (2012), 104-154.
 
\bibitem{Mikhailov-et-al} %OK
A.V. Mikhailov, A.B. Shabat, R.I. Yamilov,
The symmetry approach to the classification of
non-linear equations. Complete lists of
integrable systems,
    {\it{  Russian Math. Surveys}}  {\bf 42}, 4 (1987), 1-63.
      
\bibitem{Mikhailov-et-al2}  %OK
A.V. Mikhailov, V.S. Novikov,  J.P. Wang, 
On Classification of Integrable Nonevolutionary Equations, {\it Stud. in Appl. Math.},  {\bf 118} (2007),  419-457.

\bibitem{Miura} %OK
 R.M. Miura  Ed.
{B\"acklund Transformations, I.S.T. method and their applications},
{\it{Lecture Notes in Math.}}, Vol. 515, Springer, 1976.

\bibitem{MGK} %OK
Miura, R.M., Gardner, C.S., Kruskal, M.D.
Korteweg-de Vries equation and generalizations. II. Existence of conservation laws and constants of motion,
{\it{Journal of Math. Phys.}} {\bf 9}, 8, (1968)  1204-1209.
        
     \bibitem{walsan1} %ok 
     {W.~Oevel,  S.~Carillo},
      {Squared Eigenfunction Symmetries for Soliton  Equations: Part I,}
    {\it{J. Math. Anal. and Appl.}}
    {{\bf 217}, 1}  (1998)
     {161-178}.  \doi{ 10.1006/jmaa.1997.5707}
         
    \bibitem{walsan2} % ok
    W.~Oevel,  S.~Carillo, \ {\sl Squared
    Eigenfunction Symmetries for Soliton  Equations: Part II},
     {\it{J. Math. Anal. and Appl.}}
     {{\bf 217},  1}  (1998) 179-199.  \doi{ 10.1006/jmaa.1997.5708} 

%\bibitem{Olver} %OK
%    P.J.~Olver,
%    {Evolution equations possessing infinitely many symmetries,}
% {\it{J.~Math.~Phys.}}{\bf 18} (1977), 1212-1215.

   \bibitem{Olver:Sokolov} %OK
          P.J.~Olver and V.V.~Sokolov.
          {Integrable evolution equations on nonassociative algebras.}
 {\textit{Comm. Math. Phys.}} {\bf 193}, 245-268 (1998).

\bibitem {Rogers:1987} %ok 
C. Rogers, 
The Harry Dym equation in 2+1-dimensions: a reciprocal link with the
Kadomtsev-Petviashvili equation, {\it{  Phys. Lett.}} {\bf 120A } (1987), 15-18.

\bibitem {RogersAmes}
{C.~Rogers,  W.F.~Ames},
 {\it{Nonlinear Boundary Value Problems in Science and Engineering}}
{Academic Press}, {Boston}, {1989}.
 
\bibitem {Rogers:Carillo:1987b} %ok 
C. Rogers,  S. Carillo, $\!\!\!$ { On
reciprocal properties  of  the Caudrey-Dodd-Gib\-bon  and Kaup-Kupershmidt hierarchies}, $\!$
 {\it Phys. Scripta} {\bf 36}  (1987), 865-869.

\bibitem {RogersNucci} %OK
C. Rogers,  M.C. Nucci, On reciprocal B\"acklund transformations and the 
Korteweg- de Vries hierarchy, {\it Phys. Scr.}  {\bf 33}   (1986),  289-292.      

   \bibitem{RogersShadwick} %OK
   C.~Rogers,  W.F.~Shadwick.
      {B\"{a}cklund Transformations and Their Applications},
       {\it{Mathematics in Science and Engineering}} {Vol. 161}, 
       Academic Press, Inc., New York-London, 1982.

   \bibitem{RogersSchief} %OK
   C.~Rogers,  W.K.~Schief,
     {\it{B\"acklund and Darboux Transformations: Geometry and Modern Applications in Soliton Theory}},
      Cambridge University Press, Cambridge, 2002.
      
%\bibitem{RogersWong} C. Rogers,  P. Wong, On reciprocal Backlund transformations of inverse scattering schemes, {\textit{Phys. Scripta}} \textbf{30}, 10-14 (1984).
   \bibitem{SK}    A.K. Sawada,  A.T. Kotera, 
A method for finding N-soliton solutions of the KdV and KdV-like equations,
    {\it{J. Progr. Theor. Phys.}}  {\bf 51} (1974), 1355-1367.
    
 \bibitem{Sakh}
	\newblock A.L Sakhnovich, L.A Sakhnovich, and I. Ya.Roitberg.
	\newblock Inverse Problems and Nonlinear Evolution Equations. Solutions, Darboux Matrices and Weyl-Titchmarsh Functions.
	\newblock Studies in Mathematics Vol. 47 (De Gruyter, Berlin 2013).
	
     
%        \bibitem{Schiebold-6dic1} %ok {\red\boxed{NON }}
%        C.~Schiebold,
%     {Cauchy-type determinants and integrable systems},
% {\it{Linear Algebra and its Applications}} {\bf 433} (2010), 447­-475.
%
  \bibitem{Schiebold2009} %ok{\red\boxed{NON }}
 C.~Schiebold,
     {Noncommutative AKNS systems and multisoliton solutions to the matrix
sine-Gordon equation},
 {\it{Discr. Cont. Dyn. Systems Suppl.}} {\bf 2009} (2009), 678­-690.

%  \bibitem{Schiebold-6dic3} %ok {\red\boxed{NON }}
% C.~Schiebold,
%     {The noncommutative AKNS system: projection to matrix systems, countable
%superposition and soliton-like solutions},
% {\it{J. Phys. A}} {\bf 43} (2010), 434030.
%
%   \bibitem{Schiebold2010} %ok {\red\boxed{NON }}
%       C.~Schiebold,
%       {Structural properties of the noncommutative KdV recursion operator,}
% {\it{J. Math. Phys.}} {\bf 52} (2011), 113504.
 
 
\bibitem{Schiebold2018} 
	\newblock C. Schiebold.
	\newblock \emph{Matrix solutions for equations of the AKNS system.}
	\newblock In: "Nonlinear Systems and their Remarkable Mathematical Structures", 
		Chapter B.5, p. 256--293, Ed: N. Euler, CRC Press, Boca Raton, FL, USA 2018. 

\bibitem{russi2} %OK
S.I. Svinolupov, V.V. Sokolov,  Evolution equations with nontrivial conservative laws, {\it{Funct Anal Its Appl}}  {\bf 16} 4, (1982),  317-319. \doi{ 10.1007/BF01077866}
\bibitem{Wang} %OK
J.P. Wang, 
A List of 1 + 1 Dimensional Integrable Equations and Their Properties,
{\it{J. Nonlinear Math. Phys.}}   {\bf 9}, 1 (2002), 213-233.
 \doi{ 10.2991/jnmp.2002.9.s1.18}

\bibitem{Weiss}% {\red\boxed{NON }}
J. Weiss, 
On classes of lntegrable systems and the Painlev\'e property,
{\it{J. Math. Phys.}}   {\bf 25} (1984), 13-24.
        
%\bibitem{Wilson} {\red\boxed{NON }}
% 	G.~Wilson,
%	{On the quasi-hamiltonian formalism of the KdV equation},
%	{\it{Physics Lett. A}} {\bf 132} (1988), 445-450.      
\end{thebibliography}
\end{document}